\documentclass[12pt,a4paper]{article}

\usepackage{geometry,lipsum}
\usepackage[T1]{fontenc}
\usepackage{graphicx}
\usepackage[compress,numbers]{natbib}
\usepackage{colortbl}
\usepackage{url}
\usepackage{diagbox}
\usepackage{multirow}
\usepackage{setspace}

\title{Champuru 2: Improved scoring of alignments and a user-friendly graphical interface}
\author{Yann Sp\"ori$^1$ \\ \url{yann.spori@ulb.be}
\and Jean-François Flot$^{1,2}$ \\ \url{jean-francois.flot@ulb.be}}
\date{%
    $^1$Evolutionary Biology \& Ecology, Université libre de Bruxelles (ULB), Brussels, Belgium; tel +32 2 650 40 14; fax + 32 2 650 24 45\\%
    $^2$Interuniversity Institute of Bioinformatics in Brussels –- (IB)$^2$, Brussels, Belgium\\[2ex]%
    }
\begin{document}

\maketitle

\setstretch{2.0}

\section{Abstract}

Champuru is a web software tool that helps determine the two sequences present in mixed Sanger chromatograms obtained by sequencing simultaneously two DNA templates of unequal lengths.
A previous version (Champuru 1.0) was published as a simple Perl CGI (Common Gateway Interface) application, but the server hosting it was discontinued, which prompted us to update it and develop it further.
The new Champuru 2, implemented in Haxe and hosted at GitHub Pages, offers an improved graphical user interface as well as more sophisticated algorithms to compute alignment scores, making it more efficient at detecting the most likely alignment positions between forward and reverse traces.
Champuru 2 now make it possible to analyze other offset pairs than the one detected as most likely by the selected algorithm.
Champuru 2 is freely accessible at \url{https://eeg-ebe.github.io/Champuru/}, including both a graphical user interface (running a JavaScript version transpiled from the Haxe source code) and a compiled command-line version (obtained by transpiling the Haxe source code into C++).

\section{Keywords}
Sanger sequencing, bioinformatics, chromatogram analysis, heterozygosity, mixed traces, double peaks, transpiling, Haxe

\section{Introduction}

Although Sanger sequencing is becoming less used nowadays due to the raise of next-generation sequencing methods, it is still useful when it comes to small-scale projects.
One limitation of first-generation sequencing methods such as Sanger sequencing is that their application to mixtures of DNA sequences (e.g. resulting from sequencing nuclear markers in diploid organisms) results in double peaks, the phasing of which can be difficult~\cite{Harrigan2008,Browning2011,Flot2006}.
To help alleviate this issue, the program Champuru was published in 2007 to analyze mixed traces obtained by sequencing a mixture of two DNA sequences of uneven lengths~\cite{Flot2007}.
Here we present an update to this program, Champuru 2, with improved scoring alignments and a user-friendly graphical interface.
In contrast to Champuru 1.0 that was written in Perl with CGI (Common Gateway Interface)~\cite{Gundavaram1996}, Champuru 2 was written in Haxe~\cite{Spori2023} and is accessible
online at \url{https://eeg-ebe.github.io/Champuru}.

\section{Material and Methods}

\subsection{Input}
As with Champuru 1.0, Champuru 2 takes two strings describing the base calls of the forward and reverse chromatograms as input.
Each character of the input strings corresponds to a single or double peak in the corresponding forward or reverse chromatograms using the one-letter codes of the 1984 recommendations of the
Nomenclature Committee of the International Union of Biochemistry \cite{NomenclatureCommitteeoftheInternationalUnionofBiochemistry1985}.

\subsection{Output of the original Champuru 1.0 program}
As several improvements to the code were introduced in this new version, for reverse compatibility Champuru 2 starts by calculating the output of the previous version (1.0) of Champuru.
In case a mismatch is detected with the result of the new version, a warning message is displayed.

\subsection{Step 1 - alignment score calculation}
Like Champuru 1.0, Champuru 2 first calculates alignment scores for all possible alignments of the two input strings in order to find the best offset positions (the offset being defined as the position of the first base of the forward chromatogram by reference to the first base of the reverse chromatogram \cite{Flot2007}).
In contrast to Champuru 1.0, however, users can choose among three possible scoring schemes:
\begin{enumerate}
    \item the same alignment scoring algorithm that was already described
    in the original Champuru paper;
    \item a modification of the scoring scheme that corrects for the fact that ambiguous characters (e.g. W) can match multiple characters (e.g. A, T, and W).
Results suggests that this score calculation method works better to find the best alignments when the reconstructed consensus sequences contain a lot of ambiguities.
However, this score correction method seems to perform less well in case of short (less then 50 characters) input sequences;
    \item a scoring algorithm that returns the length of the longest stretch of consecutive matching nucleotides as a score.
    This algorithm performs well as long as the forward and reverse chromatogram sequences overlap on a long region with high-quality signal; it is not affected by the presence of low-quality basecalls in other parts of the sequences.
\end{enumerate}
In general all three matching score are very good at detecting the best offsets in the input data.
However, in cases when the signal/noise ratio is low (such as when sequences are short and/or comprise many erroneous basecalls) it can happen that one of these algorithms outperforms the others.

\begin{figure}
    \centering
    \includegraphics[width=0.7\linewidth]{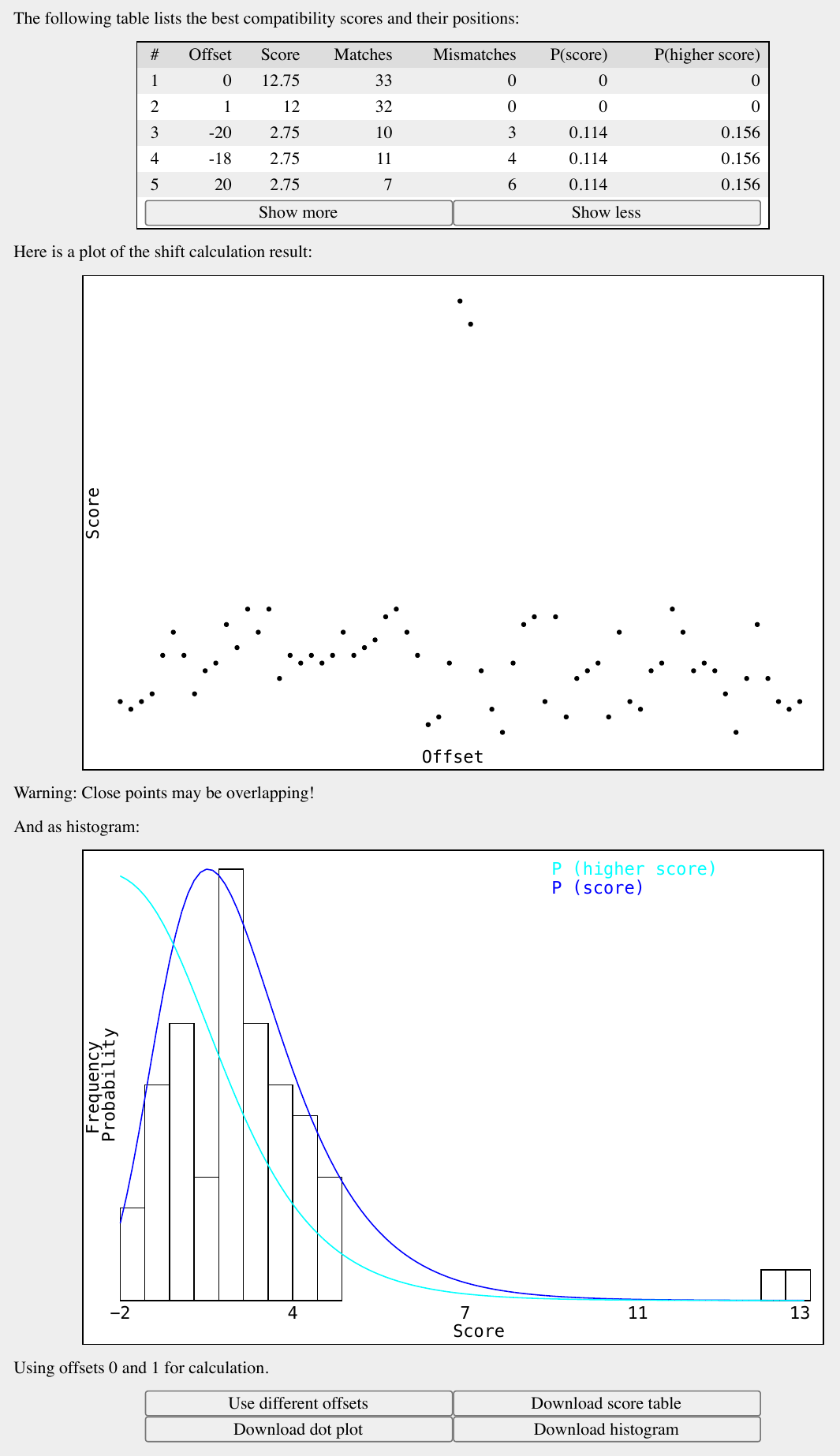}
    \caption{Graphical output of Champuru 2 for the sequence pair CTRAATTCAAATCACACTCGCGAAAWYMWKRAA (forward) and CYWRAWTYMAAWYMMMMYYSSSRAAATCATGAA (reverse) using the default scoring scheme. First, the best alignments scores are displayed in a table as well as in a scatter plot, allowing users to judge how many alignment positions to consider for downstream analyses. These alignment scores are also displayed in a scatter plot, followed by a histogram comparing the observed distribution of score with the null hypothesis of a Gumbel distribution, allowing users to detect statistically significant outliers.}
    \label{fig:Plots}
\end{figure}

Champuru 2 sorts the alignments according to their alignment scores
and - in case of equal alignment scores - according to the number of observable mismatches.
It then displays the first five best alignments in a table and produces a scatter plot of the alignment score against the offset between the two sequences (Figure \ref{fig:Plots}).
Champuru 2 also produces a histogram showing the distribution of alignment scores, with two curves superimposed on it: one showing the theoretical probability density of alignment scores under the null hypothesis of random sequences, and another curve showing the corresponding complementary cumulative distribution function (or tail function; i.e., the theoretical probability of observing an alignment with a score higher than a given value).  
To calculate these curves, it takes advantage of classical result that the probability distribution of ungapped alignment scores of two random sequences is an extreme values distribution of type I, also known as a  Gumbel distribution \cite{Karlin1990, Ortet2010}.
To estimate this null distribution, Champuru 2 shuffles 2,000 times the characters in the input sequences, each time picking up one offset randomly and calculating the corresponding alignment score.
The mean of the observed alignment scores as well as the observed standard
deviation are then fitted to a Gumbel distribution.
This allows users to visualize whether there are outliers in the alignment scores distributions, i.e. offsets for which the alignment score is much higher than expected following the null distribution.
By clicking on a dot in the scatter plot or on a box in the histogram, users can access further information about the data.
Also by moving the mouse over a line in the table, the corresponding dot in the alignment vs. offset score plot gets highlighted as well as in the histogram, allowing users to better grasp the correspondence between the three.

\begin{figure}
    \centering
    \includegraphics[width=1\linewidth]{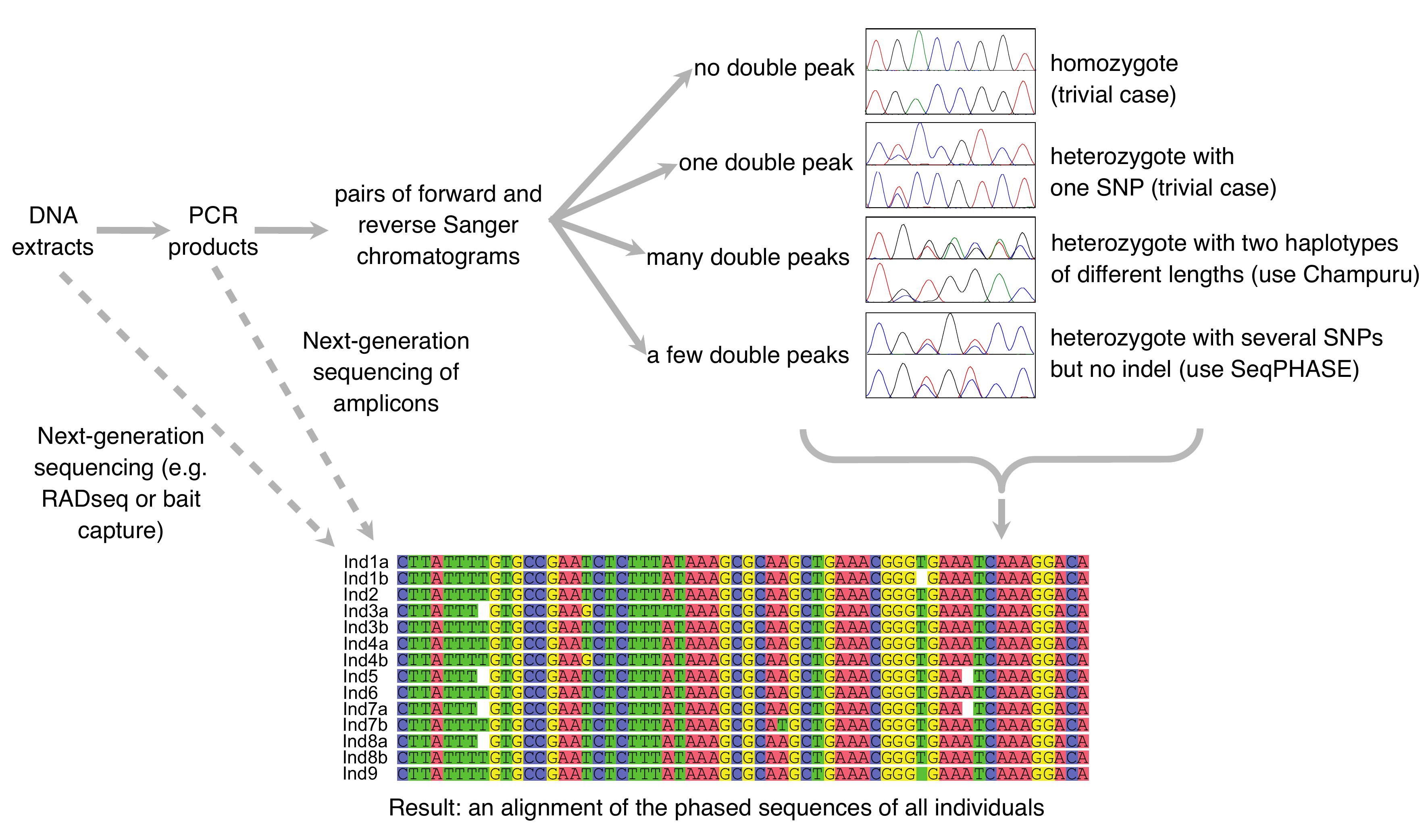}
    \caption{General flow chart showing the process for generating an alignment of phased sequences starting from Sanger chromatograms.}
    \label{fig:chart}
\end{figure}

\begin{figure}
    \centering
    \includegraphics[width=0.9\linewidth]{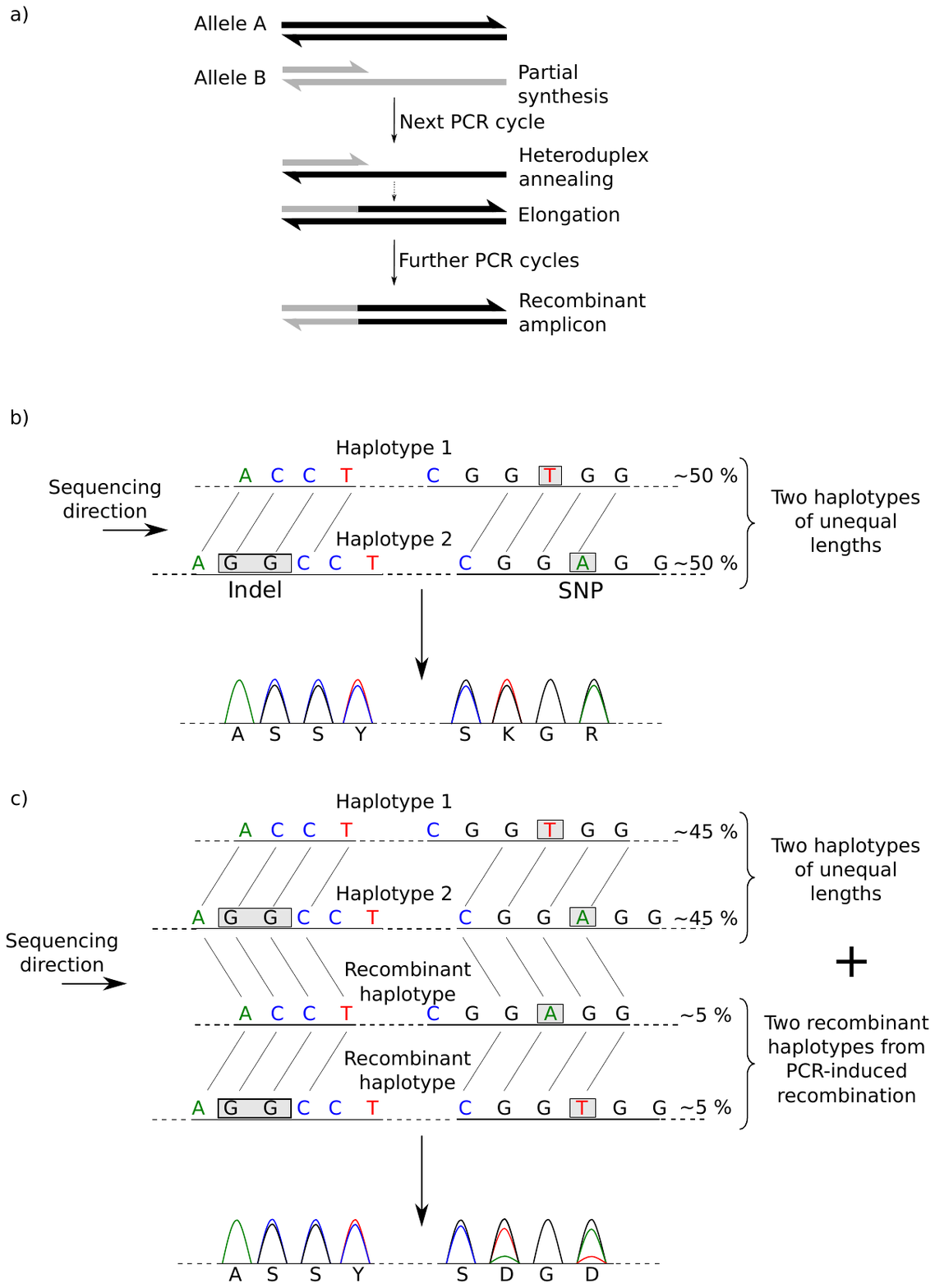}
    \caption{Schematic representation of how PCR-induced recombination between two haplotypes of different lengths can result in additional, artefactual haplotypes producing triple peaks. Modified from \cite{Flot2018}.}
    \label{fig:triplepeaks}
\end{figure}

Four possible situations can happen:
\begin{enumerate}
  \item no clear outlier is identified in the distribution of observed scores when compared to the null distribution.
  This means that the forward and reverse sequences do not align better with each other than a pair of random sequences of similar base composition, suggesting that there is a problem with the input data (for instance, the forward and reverse chromatograms do not match);
  \item a single outlier is found, indicating that there is only one way to align the forward and reverse sequences with one another.
  This suggests that the individual sequenced was homozygous at this marker (in which case any double peaks observed in the forward and reverse chromatograms are mere basecalling errors), or that it was heterozygous but with two two haplotypes of equal lengths (in which case the observed double peaks are expected to be identical between the forward and reverse chromatograms).
  For phasing heterozyogous individuals with several double peaks but no heterozygous indels, users may use the Bayesian phasing approach implemented in SeqPHASE \cite{Flot2010b, Spori2023} (Figure \ref{fig:chart});
  \item two clear outliers in the score distribution are identified, corresponding each to one possible alignment position.
  This is the expected outcome for length-variant heterozygotes, which allows reconstructing their two haplotypes with near-certainty by combining the information contained in the forward and reverse chromatograms \cite{Flot2006};
  \item more than two outliers are detected. 
  This indicates that there are more than two possible ways to align the provided forward and reverse input sequences, indicating that the amplicon contained more than two haplotypes of different lengths.
  This usually results from copy-number variation or whole-genome duplication in the starting DNA \cite{Flot2008a}, although PCR-induced recombination artefacts may sometimes produce similar patterns (Figure \ref{fig:triplepeaks}).
  In such situation, the information contained in the forward and reverse chromatograms is usually not sufficient to reconstruct the various haplotypes with certainty.
  Still, comparison of partially reconstructed haplotypes (obtained for various offset pairs) with haplotypes found in other individuals in the same population often allows to guess the most likely haplotypes present in such mixtures.
  To facilitate such guesswork, Champuru 2 now makes it possible navigate among different offset pairs: users can enter their desired offsets using the box "Use different" offsets or alternatively may rerun the analysing ticking the checkbox ``Analyze further offset pairs'', which produces at the very bottom of the page a table listing all outlier offset pairs. 
\end{enumerate}

At the end of Step 1, users can download a table containing the alignment scores and their associated probabilities.
For the following steps, Champuru 2 by default uses the two offsets corresponding to the alignments with the highest scores.

\subsection{Step 2 - consensus sequence calculation}

As a second step, Champuru calculates consensus sequences for the two input sequences in the two offsets positions determined in Step 1.
In contrast to Champuru 1.0 that only considered positions for which information from both chromatograms were available, Champuru 2 also displays nucleotides that are not aligned with the other sequence but
in small letters.

\subsection{Step 3 - sequence reconstruction}

In the simplest case when the two haplotypes differ by a single heterozygous indel and a few single nucleotide polymorphisms (SNPs), consensus sequence calculation is enough to recover the two haplotypes without ambiguities left.
However, in more complex cases some ambiguities in the consensus sequences remain to be solved, and Step 3 performs this by considering that any peak detected in the forward or reverse chromatogram has to come from one haplotype or the other.
For instance, to find out whether a Y in a consensus sequence is a C or a T, Champuru traces this Y back to matching Ys in the forward and reverse chromatograms and check whether their C or T is already explained by the other haplotype (corresponding to the other offset position).
If a Y position is a chromatogram corresponds to a C in one haplotype, then it has to be a T in the other haplotype \cite{Flot2006}.


\subsection{Step 4 - checking sequences}

Finally, the sequences reconstructed in Step 3 are checked for consistency with the chromatogram sequence data initially inputted, as well as with the result of the older Champuru v1.0 perl script.

\subsection{Step 5 - analyzing further offset pairs}

In a last step - that is only executed when the checkbox ``Analyze further offset pairs'' was ticked - Champuru 2 runs the sequence reconstruction algorithm for a number of best-scoring offset pairs.
Users can then click in the box of any specific offset pair to obtain the corresponding results.


As this calculation takes extra time, this option is not checked by default.
\section{Discussion}

Compared to the previously published version \cite{Flot2007}, this improved version offers the following enhancements:
\begin{enumerate}
  \item several scoring schemes for finding the best alignment positions;
  \item a series of graphical outputs highlighting the best alignment positions and their respective scores (only in the graphical version of the program);
  \item a calculation of the statistical significance of the best scores obtained compared to the null hypothesis of a Gumbel distribution;
  \item the possibility of choosing alternative pairs of offsets (instead of solely the best-scoring offset pair).
\end{enumerate}

Our reimplementation of Champuru using the Haxe language not only ensures that this useful piece of code is again available to everyone, but also that it is available both as a user-friendly graphical version and as a command-line version.
Being hosted on GitHub Pages (\url{https://eeg-ebe.github.io/Champuru}), the code runs directly in the web browser of the user on their own computer, eliminating the hurdle of maintaining an online, secure server and also ensuring that users' sequence data are not transmitted over the internet.
Compiled C++ versions for linux and MacOS are provided for download (\url{https://eeg-ebe.github.io/Champuru/download.html}) but since the original Haxe code is available at \url{https://github.com/eeg-ebe/Champuru}, interested users may also transpile it into a variety of other languages such as Neko, Python etc. if they wish.



\newpage
\bibliographystyle{unsrtnat}

\bibliography{mybib}

\end{document}